\documentclass{aa}
\usepackage{hyperref}
\pdfoutput=1 
\usepackage{natbib}
\usepackage{txfonts}
\usepackage{graphicx}
\usepackage{indentfirst}
\usepackage{booktabs}

\begin{document} 

\title{Evolution of the Alfvén Mach number associated with a coronal mass ejection shock}
\titlerunning{The evolution of shock Alfvén Mach number}
\author{Ciara A. Maguire \inst{1,}\inst{2}     \and    Eoin P. Carley   \inst{1,}\inst{2} \and Joseph McCauley\inst{1} \and Peter T. Gallagher \inst{2,}\inst{1} }

\institute{School of Physics, Trinity College Dublin, Dublin 2, Ireland.
\and School of Cosmic Physics, Dublin Institute for Advanced Studies, Dublin D02 XF85, Ireland. \\
\email{cmaguir4@tcd.ie}
}

\date{Received August 2, 2019; accepted November 29, 2019}
\abstract
{The Sun regularly produces large-scale eruptive events, such as coronal mass ejections (CMEs) that can drive shock waves through the solar corona. Such shocks can result in electron acceleration and subsequent radio emission in the form of a type II radio burst. However, the early-phase evolution of shock properties and its relationship to type II burst evolution is still subject to investigation. Here we study the evolution of a CME-driven shock by comparing three commonly used methods of calculating the Alfvén Mach number ($M_A$), namely: shock geometry, a comparison of CME speed to a model of the coronal Alfvén speed, and the type II band-splitting method. We applied the three methods to the 2017 September 2 event, focusing on the shock wave observed in extreme ultraviolet (EUV) by the Solar Ultraviolet Imager (SUVI) on board GOES-16, in white-light by the Large Angle and Spectrometric Coronagraph (LASCO) on board SOHO, and the type II radio burst observed by the Irish Low Frequency Array (I-LOFAR). We show that the three different methods of estimating shock $M_A$ yield consistent results and provide a means of relating shock property evolution to the type II emission duration. The type II radio emission  emerged from near the nose of the CME when $M_A$ was in the range 1.4-2.4 at a heliocentric distance of $\sim$1.6 $R_\odot$. The emission ceased when the CME nose reached $\sim$2.4 $R_\odot$, despite an increasing Alfvén Mach number (up to 4). We suggest the radio emission cessation is due to the lack of quasi-perpendicular geometry at this altitude, which inhibits efficient electron acceleration and subsequent radio emission.}

\maketitle

\section{Introduction}
Coronal mass ejections (CMEs) are large eruptions of plasma and magnetic field that propagate from the low solar corona into the heliosphere. If a CME reaches a speed that exceeds the local background Alfvén speed, a plasma shock forms, most commonly at the CME nose and flanks \citep{Cho2007,Carley2013,Zucca2014}. CME-driven shock properties, such as Alfvén Mach number ($M_A$), have been calculated from a variety of observational methods in the past. However, these methods have rarely been compared directly and, hence, their reliability is currently unknown. Here we compare three commonly used methods to derive $M_A$, providing a measure of their consistency and also providing insight into the evolution of shock and radio emission characteristics in the early phases of CME eruption.
\newline \indent
CME-driven shock signatures can be observed at a variety of wavelengths, most predominantly in extreme ultraviolet (EUV), white-light, and radio \citep{Grechnev2011a,Vourlidas2012,Mancuso2019}. Each wavelength range offers an independent and unique method of determining $M_A$. In EUV and white-light, the geometry of both driver and shock can be derived from images, providing a measure of various shock properties, such as compression ratio and Mach number. This method has its origin in laboratory experiments that were designed to study shock formation around various types of blunt obstacles \citep{Seiff1962, Spreiter1966}. The theory was later developed for Earth’s magnetospheric bow shock \citep{Farris1991}, with \cite{RussellMulligan2002} subsequently adapting the theory to understand the relationship between interplanetary CMEs and the geometry of their associated bow shock. This has recently been applied to CME-driven shocks imaged in white-light, showing $M_A$ to be in the range of 3 to 5 at heliocentric distances up to 0.5 AU \citep{Maloney2011,GopalswamyYash2011,Poomvises2012}. The method has also been used for EUV images of CME and shocks early in their evolution \citep{Gopalswamy2011b}, showing that at a heliocentric distance of $<$1.5 $R_\odot$, values of $M_A$ are typically in the range of 1.5 to 3.7. 
\newline \indent
A different approach that makes use of EUV and white-light observations is to compare CME speed (derived from images) to a data-driven model of the coronal Alfvén speed \citep{Zucca2014}. This method has recently been modified to produce measures of $M_A$ in 3D coronal environments, showing values of 1 to 3 at heliocentric distances of $<$2 $R_\odot$
\citep{Rouillard_2017,Zucca2018,Morosan2019}. Some studies have used EUV observations to derive coronal plasma properties by taking the ratio of filter bands in EUV telescopes \citep{ Berger2012}. Differential emission measure analysis has also been used to estimate changes in temperature and density in the shock sheath to infer compression ratios and $M_A$ values \citep{Kozarev2011,Kozarev2015,Frassati2019}.
\begin{figure*}[!ht] 
\centerline{\includegraphics[width=14.5cm ,trim=1cm 2cm 1cm 4cm]{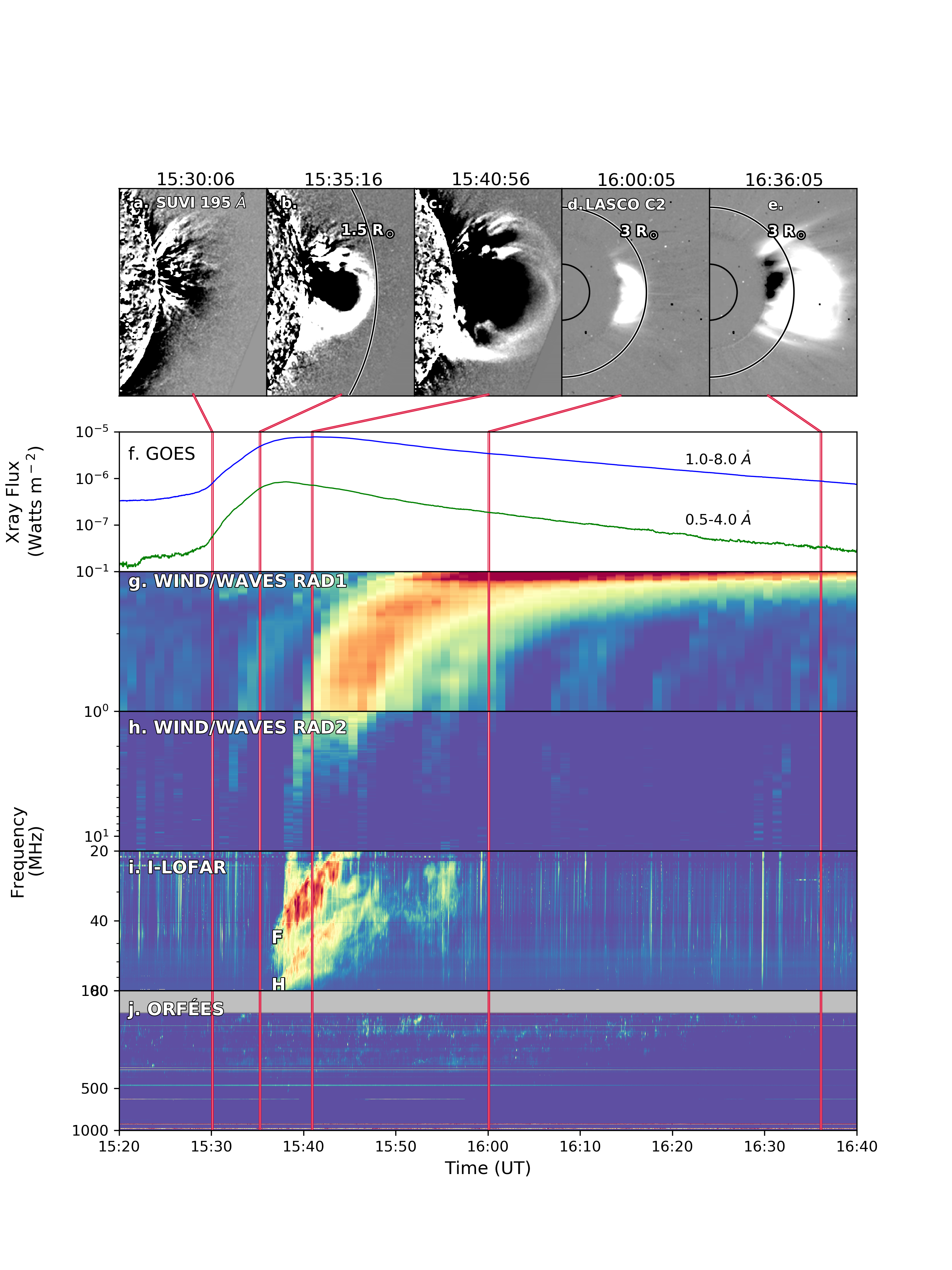} }
\caption[EUV, white-light, Dynamic radio spectra and light-curves of radio, soft X-ray flux profiles during the 2017 September 2 C7.7 class solar flare]{Base difference images of the CME observed with SUVI 195 \r{A} (a-c) and LASCO C2 (d $\&$ e). (f) GOES 0.5-4 \r{A} and 1-8 \r{A} soft X-ray flux of the C7.7 class solar flare. The central panels plot the radio dynamic spectra from the event covering a frequency range from 0.5-1000 MHz; (g) WIND/WAVES RAD 1 (20–1040 kHz), (h) WIND/WAVES RAD 2 (0.5-16 MHz), (i) I-LOFAR (20-88 MHz), (j) ORFÉES (140-1000 MHz). The I-LOFAR dynamic spectrum shows a type II radio burst with fundamental (F) and harmonic (H) components. }
\label{figure1}
\end{figure*}

\begin{figure*}[!ht]  
\centerline{\includegraphics[width=18.5cm,trim=0.5cm 0.5cm 0.9cm 0.5cm]{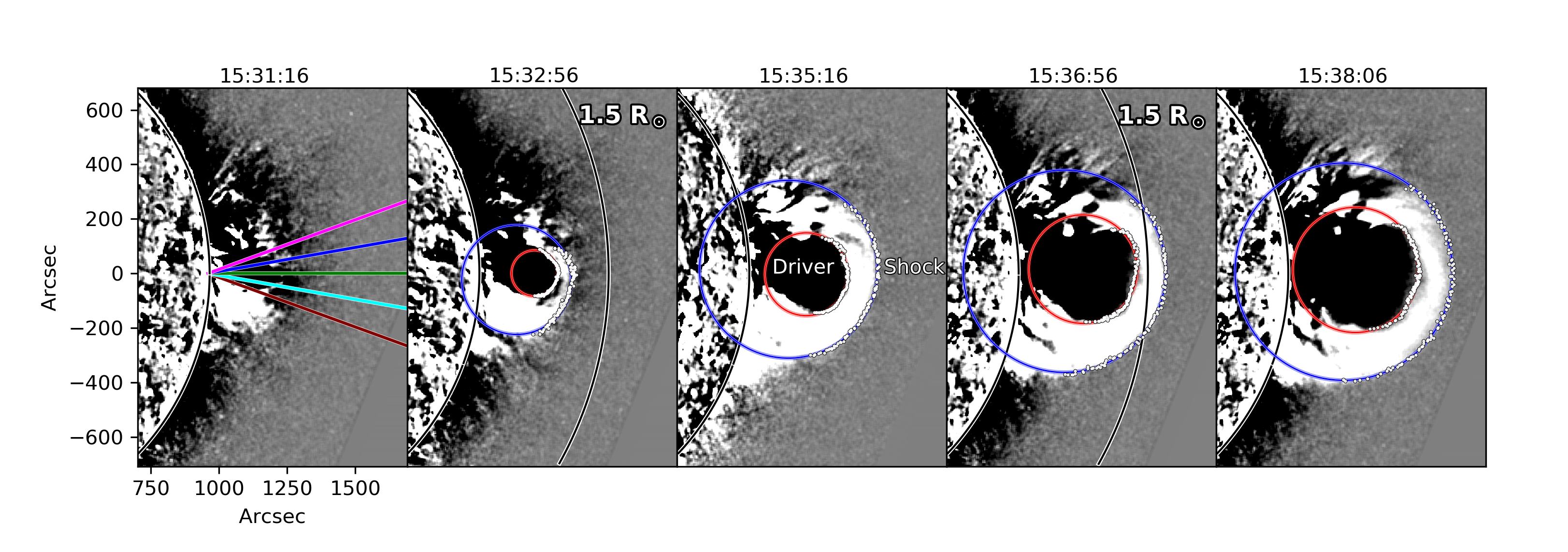}
}
\caption[Base-difference images of the flaring region from GOES/SUVI ]{Base-difference images of the flaring region on the western limb of the Sun made by SUVI in the 195 \r{A} channel from 15:31:16  to 15:38:06 UT, showing a dark region (interpreted as the driver) surrounded by an intensity enhancement (interpreted to be the shock sheath). The image contrast range was reduced to enhance these features. The SUVI base image is the average of 5 images prior to the start of the flare. The five traces examined in order to determine the CME apex are shown in the first image. The large red and blue circles indicate the fitting to the edges of the eruptive plasma and shock front respectively, with the circle width representing a $\pm$1 $\sigma$ ($\pm$6") uncertainty. The dots are the points along the driver and shock front chosen using a simple point-and-click method.}  

\label{figure2}
\end{figure*}
\indent 
While EUV and white-light images provide an indirect measure of shock properties from their geometry and kinematics, radio observations can be used to probe plasma shock properties more directly. At radio wavelengths, we observe type II radio bursts as evidence for shocks, often with two emission bands corresponding to the fundamental and first harmonic of the local plasma frequency \citep{Wild1950,Wild1962}. The emission bands sometimes split into two thinner subbands with similar morphology and intensity variations, a phenomenon known as band-splitting \citep{Nelson1985}. These subbands can often appear as distinct separate bands, fragmented subbands \citep{Chrysaphi2018,Mahrous2018} or a single emission band with a large bandwidth \citep{Mann1995}. This phenomenon is said to be a consequence of simultaneous emission from the plasma in front of and behind the shock, referred to as the upstream-downstream theory \citep{Smerd1974}. Applying the Rankine Hugoniot jump conditions and using the relative bandwidth of the band-split, we can derive the shock compression ratio and estimate $M_A$, with several authors calculating values in the range of 1.3 to 1.6 at heliocentric distances of $\sim$1.2 to 2 $R_\odot$ \citep{Vrsnak_Mag2002,Zimovets2012,Zucca2018}. This method is in question however, as the precise nature of the band-splitting is still under debate \citep{Du2015}. 
\newline \indent 
To date, a variety of methods have been developed to estimate shock characteristics from different wavelength observations, but they have rarely been compared \citep{Ma2011,Kozarev2011,Kouloumvakos2014}. Here we compare three commonly used methods to derive $M_A$ namely: shock geometry in EUV images, a comparison of CME speed to a data-driven model of Alfvén speed, and the type II band-splitting method. This allows us to test the consistency of the methods, but it also allows us to derive more detailed shock characteristics than would normally be available using just one method. We determine the location of the type II radio emission along the CME front and we relate the change in the angle between the local shock norm and coronal magnetic field direction to the onset and ceasing of the type II radio emission. In Section \ref{Observations}, the observations of a specific CME and type II radio burst are presented. The three methods to determine shock characteristics and evaluate $M_A$ are explained in Section \ref{method}. In Section \ref{result}, we compare the results from the three methods and conclusions are discussed in Section \ref{conclusions}.

\begin{figure*}[ht!]   
\begin{center}
\includegraphics[width=16cm, trim=0cm 0cm 0cm 0cm]{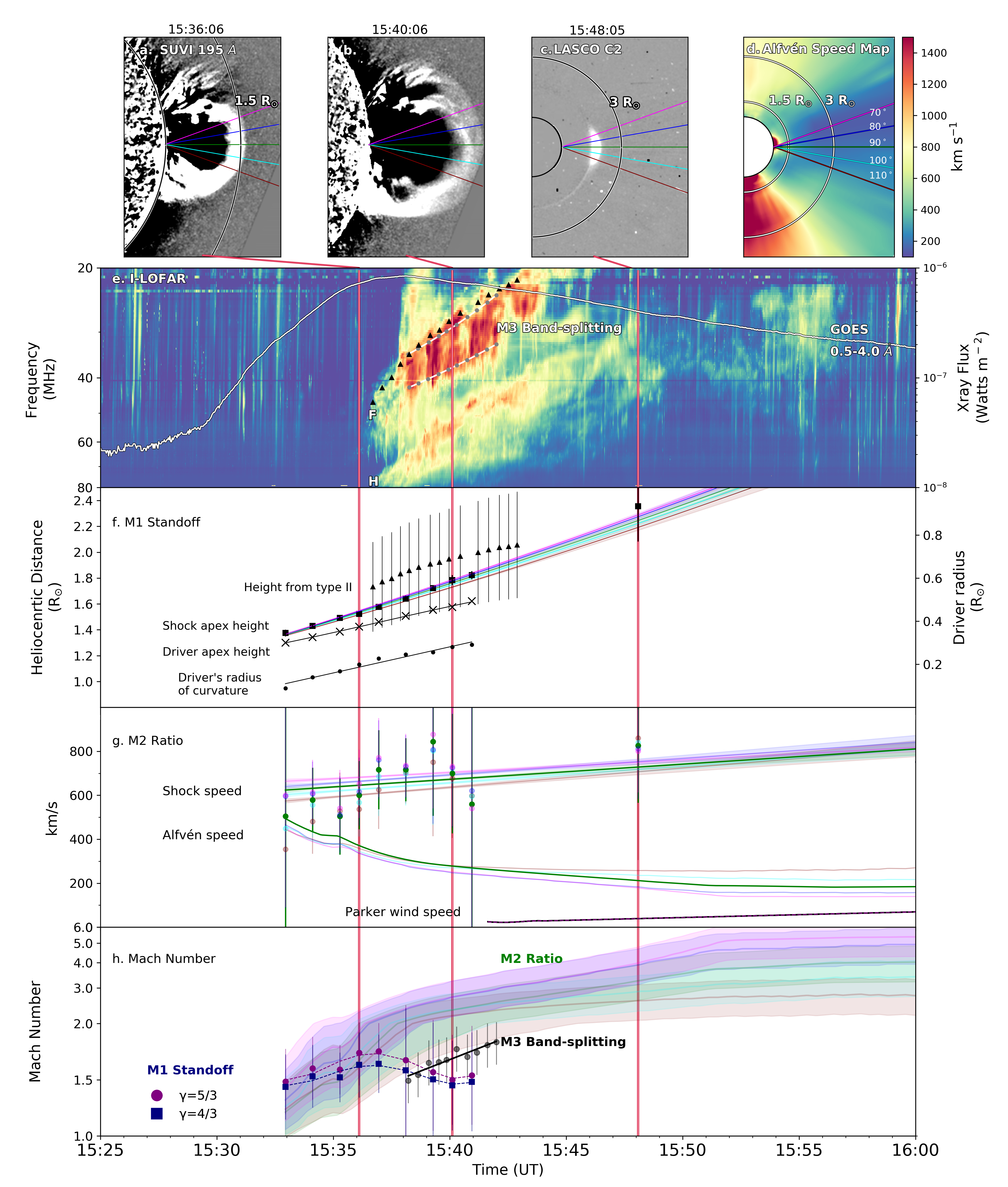}
\end{center}
\caption{
(a-c) Base difference images from SUVI 195 \r{A} and LASCO C2. (d) The 2D Alfvén speed map produced from the model described in \citet{Zucca2014}. Overlaid are the color-coded traces used in the ratio method, starting at 70$^\circ$ to the solar north and separated by 10$^{\circ}$. (e) Type II radio burst dynamic spectra from I-LOFAR showing fundamental (F) and harmonic (H) components. The grey points indicate the upper and lower edges of the fundamental component. The black triangles on the fundamental mark the points used in the height of radio emission calculations. Overlaid is GOES high energy (0.5-4 \r{A}) X-ray flux. (f) The triangles indicate the source height of the type II fundamental derived from the 2D electron density map. The squares mark the shock front height along the green trace from  SUVI and LASCO C2 measurements. The height-time profiles of the shock front from each trace, with the colors corresponding to the traces indicated in panel a-c. The CME driver height along the green trace is marked by crosses and the radius of curvature of the CME is marked by points. (g) The estimated shock speed with uncertainties and Alfvén speed along the five traces marked in the images in the top panel. The black line plots the Parker solar wind solution from \cite{Mann2002}. (h) $M_A$ evaluated using three methods: M1 (standoff distance), M2 (CME speed to Alfvén speed ratio) and M3 (band-splitting).}
\label{figure3}
\end{figure*}

\section{Observations}
\label{Observations}
\indent
A GOES C7.7 class flare (Figure \ref{figure1}f) began on 2017 September 2 at $\sim$15:23 UT from the active region (AR) NOAA 12672 (N05W90). The flare was associated with a CME that was observed at EUV wavelengths by GOES Solar Ultraviolet Imager \citep[SUVI;][]{Seaton2018} and in white-light by the Large Angle and Spectrometric Coronagraph C2 \cite[LASCO;][]{Brueckner1995}. The CME propagated with an average velocity of $\sim$710 km s$^{-1}$, derived from a linear fit to the height-time measurements from SUVI and LASCO C2. Both flare and CME occurred on the western solar limb providing a plane-of-sky (POS) view of the eruption, as shown in Figure \ref{figure1} (a) to (e). \newline \indent In Figure \ref{figure1} (g) to (j) the spectral radio observations from various ground-and-space instruments are shown, namely, WIND WAVES spectrographs RAD1 and RAD2 \citep{Bougeret1995} observing between 20-1040 kHz and 1.075-13.825 MHz, respectively (g $\&$ h); Irish Low Frequency Array (I-LOFAR) observing between 10-240 MHz (i); and the radio spectrograph, Observation Radio Frequence pour l’Étude  des  Eruptions  Solaires (ORFÉES), observing between 140-1000 MHz (j). Panel (i) shows the type II burst observed by I-LOFAR at $\sim$15:36 UT with relatively well defined fundamental and first harmonic components, indicated by F and H in the figure. The center of the fundamental and harmonic components were first observed at $\sim$35 and $\sim$75 MHz respectively. The fundamental drifted gradually with a mean rate of -0.05 MHz s$^{-1}$. Superimposed over the type II is a type III radio burst that extends from $\sim$40 to $\sim$1 MHz observed at $\sim$15:37 UT. There is no significant emission above 100 MHZ, suggesting no radio emission escaped from low altitudes in the corona.

\section{Data Analysis}
\label{method}
In the following, we determine the CME-driven shock characteristics via three commonly used methods. This allows us to compare these methods and determine if the results given by each are consistent. The variety of shock characteristics provided by each method also allows us to determine the relationship between the eruptive structure seen in EUV and the type II burst as observed in radio. Specifically, we determine where the radio burst was generated in relation to the eruptive structure (nose or flank) and the kind of coronal environment that lead to shock-accelerated electrons and subsequent radio emission.
\subsection{Method 1: Standoff Distance }
\indent It is possible to derive shock properties from its geometry in images. \cite{RussellMulligan2002} proposed that $M_A$ is related to the normalized standoff distance ($\delta$), that is, the ratio of the standoff distance ($\Delta$) to the radius of curvature of the CME ($R{_c}$), by 
\begin{equation}
M_A=\sqrt{1 + [1.24\delta -(\gamma-1)/(\gamma+1)]^{-1}}
\label{eq:mach}
\end{equation}

where $\gamma$ is the adiabatic index ($\gamma$=5/3 for an ideal mono-atomic gas and $\gamma$=4/3 for ideal mono-atomic relativistic gas). We applied this method to base difference images from GOES/SUVI in which we clearly see the evolution of the eruption on the western limb of the Sun. In Figure \ref{figure2}, the brighter region is interpreted as the coronal plasma compressed by the transit of the shock wave driven by the CME, a region often referred to as the shock sheath. The darker circular feature that propagates away from the flaring region is identified as the CME that drives the shock. In ten trials, ten points were manually chosen along the driver and shock fronts, marked as white dots. The driver and shock fronts were subsequently fit with a circle, indicated by the red and blue overlying circles with the width of the circle representing a $\pm$1 $\sigma$ uncertainty. We considered the five traces (as seen in the first panel of Figure \ref{figure2}) and determined where each trace intersected the blue circle to obtain a height-time profile along each trace. The height-time profile associated with the green trace was found to be at the largest height and was therefore taken to be the apex of the CME. Under this assumption, the standoff distance ($\Delta$) was taken to be the distance between the nose of the CME driver and shock front along the green trace. The standoff distance and radius of curvature of the CME ($R{_c}$) were evaluated for nine instances from 15:32:56 to 15:40:56 UT. Over this time frame the LE of the shock travelled from a heliocentric distance of $\sim$1.4 to $\sim$1.9 $R_{\odot}$ as shown in Figure \ref{figure3}(f). For each observation $M_A$ was calculated using Equation \ref{eq:mach}, the results of which are shown in Figure \ref{figure3}(h) and discussed in Section \ref{compare}.

\subsection{Method 2: CME Speed to Alfvén Speed Ratio}
\indent $M_A$ was calculated by taking the ratio of the CME speed ($v_{CME}$) to the local background Alfvén speed ($v_{A}$). Propagation speed for the CME was derived from EUV and white light imaging while the Alfvén speed was derived from the \cite{Zucca2014} model. A height-time profile of the CME was derived from SUVI and LASCO C2 base difference images using a point-and-click technique to track the CME’s leading edge above the solar limb.  As it is possible that the shock formed over an extended region around the nose we examined five traces around the shock nose marked and color-coded in Figure \ref{figure3}(a-d).  The traces originate at the active region from the solar limb, starting at 70$^{\circ}$ to the solar north and are separated by 10$^\circ$. The height-time profiles along each trace were fitted using a second order polynomial, shown in Figure \ref{figure3}(f). The faint bands represent the uncertainty in position  at all points in time ($\pm1$ $\sigma$), which was determined from the fit. The derivative of the height-time fits gave continuous velocity profiles, shown in Figure \ref{figure3}(g) and the velocity uncertainties were propagated from the position uncertainties. The motivation for using a parametric fit was to obtain smooth velocity profiles, as the velocities obtained by taking the first numerical derivative of the height measurements have a large scatter as shown in Figure \ref{figure3}(g).
\newline \indent A section of the 2D plane-of-sky Alfvén speed map on the day of the event produced from the method described in \citet{Zucca2014} model is shown in Figure \ref{figure3}(d). Alfvén speeds reached $\sim$10$^3$ km s$^{-1}$ around the active region and decreased to $\sim$200 km s$^{-1}$ at higher altitudes. The Alfvén speed along the five traces marked in Figure \ref{figure3}(d) were extracted from the map. 
\newline \indent When calculating $M_A$ for each trace, the speed of the CME relative the solar wind speed is used such that $M_A$ is $(v_{CME}$-$v_{wind})/v_{A}$ where $v_{wind}$ is taken as the Parker solution to the solar wind, as in \citet{Mann2002}. The results for $M_A$ using this ratio method are shown in Figure \ref{figure3}(h) and discussed in Section \ref{compare}.

\subsection{Method 3: Band-splitting}
The type II observed by I-LOFAR shown in Figure \ref{figure3}(e), does not present "classical" band-splitting in the form of distinctive split bands, instead we observe emission bands with large bandwidth. Similar observations presented by \cite{Mann1995} suggest that the relative instantaneous bandwidth of the type II can be used to infer the density jump across the shock wave and $M_A$ values. We tested the validity of using the relative instantaneous bandwidth to derive $M_A$ values, assuming the \cite{Smerd1974} upstream-downstream theory. Points were selected along the lower and upper boundaries of the fundamental, marked by grey dots in Figure \ref{figure3}(e). The obtained points were then used to derive the relative instantaneous bandwidth and measure the compression ratio. To determine $M_A$ values, we used the expression from \citet{Vrsnak_Mag2002} for a quasi-perpendicular shock:
\begin{equation}\label{eq:Ma}
M_A = \sqrt{\frac{X(X + 5 +5\beta)}{2(4-X)}}
\end{equation}
where $X$ is the compression ratio and $\beta$ is the plasma-to-magnetic pressure ratio. From EUV imaging we determined the shock to be at a height of $\sim$0.6 $R_{\odot}$ above an active region, where we estimate $\beta$ $\gtrsim$0.2 from \cite{Gary2001}. Values of $M_A$ derived from the bandwidth of the fundamental are shown in Figure \ref{figure3}(h) and discussed in Section \ref{compare}.

\section{Results}
\label{result}
\subsection{Comparison of three methods}
\label{compare}
\indent We employed three commonly used methods to derive $M_A$ associated with the CME-driven shock produced from a C7.7 class flare on 2017 September 2. The first method involved measuring the standoff distance at the shock nose and radius of curvature of the CME from EUV images taken by SUVI. We tracked the CME from a heliocentric distance of 1.2 to 1.9 $R_{\odot}$, until both the shock and CME expanded outside the instrument's field of view. The shock-normalized standoff distance ($\delta$) was found to be approximately constant with a mean of 0.7$\pm$0.2. For an adiabatic index of 5/3, $M_A$ steadily increased from 1.5$\pm$0.3 to 1.8$\pm$0.3 before decreasing steadily again to 1.5$\pm$0.1. An adiabatic index of 4/3 showed similar behavior as seen in Figure \ref{figure3}(h), with results comparable to \cite{Gopalswamy2011b}. 
\newline \indent The second method involved deriving the CME speed and Alfvén speed and taking the ratio to evaluate $M_A$ (whilst accounting for the solar wind). As it is possible that the shock formed over an extended region around the nose, we examined five traces over this region. We found the Alfvén speed decreased from $\sim$500 km s$^{-1}$ at a heliocentric distance of $\sim$1.4 $R_{\odot}$ to $\sim$200 km s$^{-1}$ at $\geq$2 $R_{\odot}$. Using $(v_{CME}$-$v_{wind})/v_{A}$, we found that on average $M_A$ increased steadily from $\sim$1.5 up to $\sim$4 over a time frame of $\sim$17 minutes, which is in agreement with results from \cite{Zucca2018} and \citep{Morosan2019}.
\newline \indent The third method, which used measurements of the relative instantaneous bandwidth in the type II fundamental component found $M_A$ values to lie in the range of 1.5 to 1.8, which is in agreement with previous studies \citep{Vrsnak2001,Zucca2014b,Chrysaphi2018}.

\begin{figure*}[!ht]   
\begin{center}
\includegraphics[width=17cm ,trim=0cm 0.5cm 0cm 0cm]{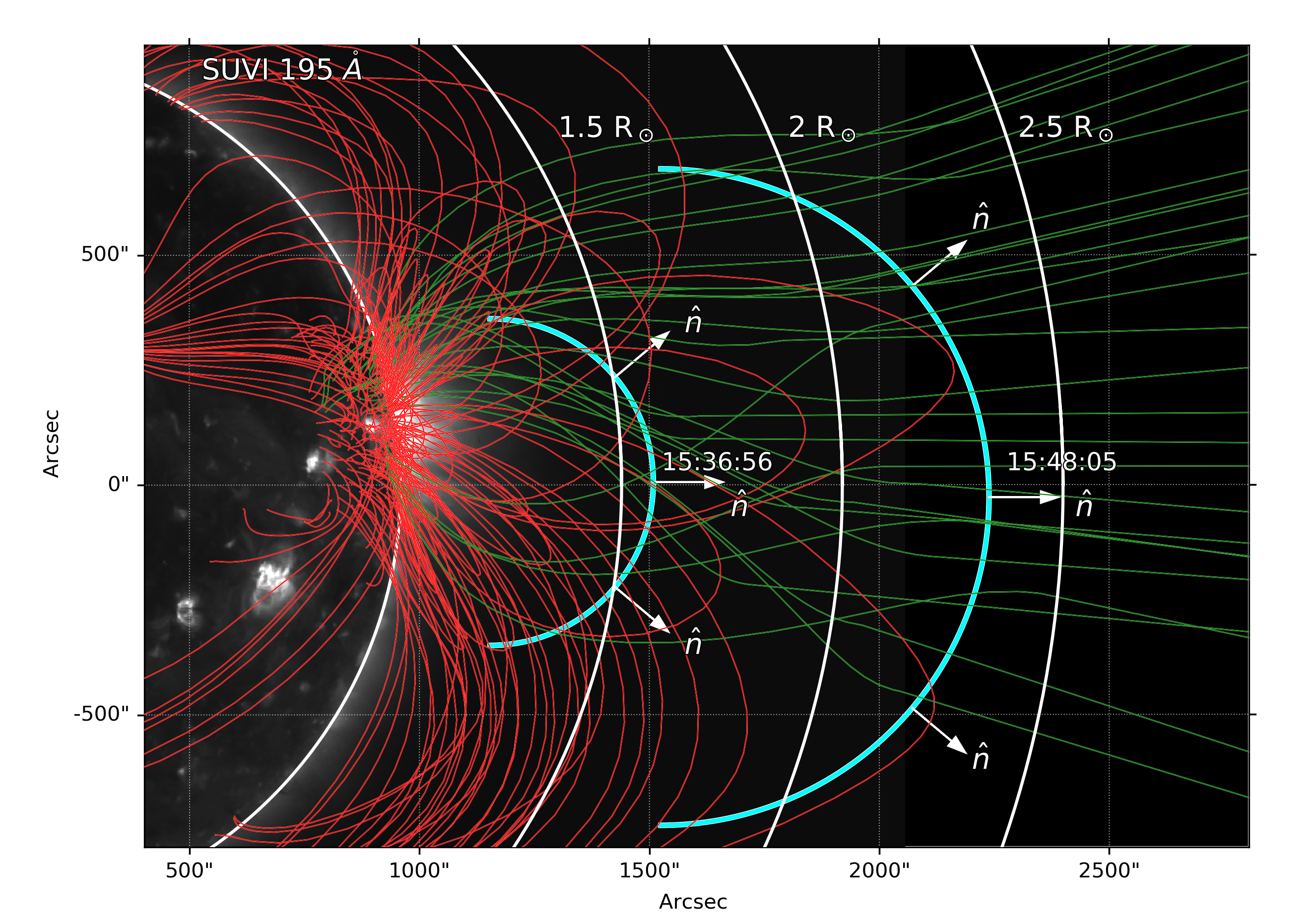}
\end{center}
\caption[PFSS]{
Potential magnetic field lines around the region of interest on 2017 September 2 overlaid on a GOES/SUVI image. A selection of closed (red) and open (green) field lines are shown. The cyan arcs represent the position of CME's leading edge when the type II emission began at 15:36:56 UT and ceased 15:48:05 UT. The white arrows indicate the direction of the shock norm. The shock geometry is more likely to be quasi-perpendicular at the type II onset while a quasi-parallel configuration is more probable when radio emission ceases.}
\label{figure4}
\end{figure*}

The results from all three methods were initially similar ($\sim$1.5) but diverged later on. The inherent uncertainties associated with each method may explain the discrepancy between the results. $M_A$ values from the standoff method deviate after $\sim$15:37 UT, which may be due to the CME front leaving the field of view at this time, making the $M_A$ more difficult to determine and less reliable. In addition, after $\sim$15:39 UT, $M_A$ values from the ratio method were slightly larger (>2) than those derived from the band-splitting method (1.5 to 1.8), which may be a consequence of the uncertainties that exist in deriving $v_{CME}$ from imaging, $v_{A}$ from the \cite{Zucca2014} model (given that the model is based on a combination of electron density models and a Potential Field Source Surface (PFSS) Model) and $v_{wind}$ from the \cite{Mann2002} model. Furthermore, the band-splitting method is also model dependent and assumes a quasi-perpendicular shock and $\beta$ $\gtrsim$0.2, which may not always be the case. \newline \indent
It is important that future studies of coronal shock properties err on the side of caution and use more than one method to determine values of $M_A$. In this study, we considered results from all three methods and look at the general trend, which suggests when the shock was at a heliocentric distance of 1.4 $R_{\odot}$, $M_A$ was $\sim$1.5 and increased up to $\sim$4 as the shock propagated to $\sim$2.5 $R_{\odot}$.

\subsection{Comparison of shock EUV kinematics with type II kinematics to determine radio source location}
\label{radioheight}
In order to determine where the radio burst was generated with respect to the eruptive feature we use the type II and an electron density model to calculate the source height of the emission. Points were extracted from the fundamental component, marked by black triangles in Figure \ref{figure3}(e) and the electron density ($n_e$) corresponding to each frequency ($f_p$) was calculated using the standard relationship 
\begin{equation}
f_p = 8980\sqrt{n_e}
\end{equation} 
where $n_e$  is expressed in cm$^{-3}$ and $f_p$ is in MHz. Using the electron density map produced by the \citet{Zucca2014} model, we found that the type II propagated from a heliocentric distance of $\sim$1.8 to $\sim$2 $R_{\odot}$ with uncertainties in height of $\sim$20$\%$ \citep{Zucca2014b}, as shown in Figure \ref{figure3}(f). 
A comparison with shock EUV kinematics suggest the type II was at a higher altitude than the CME apex and therefore likely located around the nose of the CME (as opposed to the flanks). This tells us what shock geometry existed that lead to shock-accelerated electrons and subsequent radio emission, which is in agreement with previous studies \citep{Carley2013,Zucca2014}.

\section{Discussion}
\subsection{Why does the type II emission start?}
As seen in Figure \ref{figure3}(h), at 15:33 UT, $M_A$ is greater than unity according to Methods 1 and 3, which would imply shock formation yet type II emission does not start until 15:37 UT when $M_A$ $\sim$1.7. In order to investigate why this is the case we study the relationship between the shock geometry with respect to the magnetic field and the associated type II. 
This requires a calculation of the magnetic field prior to the eruption using a PFSS model \citep{Stansby2019}. No flares occurred in the days prior to the event, meaning the magnetic field did not change significantly and we can assume the PFSS model is reasonably reliable. The PFSS model overplotted on a SUVI 195 \r{A} image is shown in Figure \ref{figure4}, with closed magnetic field lines in red and the open magnetic field lines in green. The two cyan arcs represent the location of the CME nose at the onset and cessation of the type II burst. At $\sim$15:36 UT, the CME was at a heliocentric distance of $\sim$1.6 $R_\odot$ and passed through a region with numerous closed magnetic field lines.
\newline \indent The angle $\theta_{Bn}$ between the local shock norm $\hat{n}$ and upstream magnetic field direction $\hat{B}$  is an important quantity in deciding what electron acceleration mechanism occurs. In the quasi-perpendicular case ($\theta_{Bn}$ $\geq$ 45$^\circ$) electron reflection and acceleration at the shock is more likely to occur via the shock drift acceleration (SDA) mechanism \citep{Holman1983,Street1994, Schmidt2012}. In this mechanism, the charged particles experience a grad-B drift along the shock front and gain energy due to the induced electric field, a result of
\textbf{\vec{v} $\times$ \vec{B}} flow at the shock boundary \citep{Hoffmann1950}. The accelerated electrons result in type II radio emission through the plasma emission mechanism \citep{Melrose1975,Ginzburg1993}. As seen in Figure \ref{figure4}, during the initial stages of the eruption, the shock-to-field geometry was mostly quasi-perpendicular, which suggests conditions were favorable for SDA. Despite favorable shock geometry and a $M_A$ greater than unity, the type II only forms when the shock reaches a heliocentric distance of $\sim$1.6 $R_\odot$ and $M_A$ was in the range 1.4-2.4. The onset of the type II may be explained when the shock $M_A$ becomes supercritical, that is, shock starts accelerating particles. The supercritical $M_A$ number is greater than unity and depends on various shock parameters, including upstream $\beta$ and shock angle $\theta_{Bn}$. \cite{Edmiston1984} showed that the supercritical Mach number is $\lesssim$2.4 for a quasi-perpendicular shock with $\gamma$=5/3 and $\beta$ $\gtrsim$0.2. This is in agreement with the value of $M_A$ determined at the start time of the radio emission in our work, which is in the range 1.4-2.4 (see Figure \ref{figure3}h). This may explain why the type II emission was not observed despite a $M_A$ greater than unity, that is, a supercritical $M_A$ was required before particle acceleration began and radio emission was generated.

\subsection{Why does the type II emission stop?}
At $\sim$15:48 UT, the CME nose reached a heliocentric distance of $\sim$2.4 $R_{\odot}$ and the type II emission ceased despite the shock being super-Alfvénic and a $M_A$ greater than 4, as seen in Figure \ref{figure3}(h). It is possible that the radio emission continued but it was absorbed before it could not escape, however, this is unlikely for the harmonic emission. Alternatively, the cessation of radio emission could imply the conditions were no longer favorable for electron acceleration. This may be a result of a change in shock geometry with respect to the magnetic field. As seen in Figure \ref{figure4}, at the time of the type II onset the shock nose propagated through a region where the geometry is mostly quasi-perpendicular. Several minutes later at $\sim$15:48 UT, the CME nose reached a heliocentric distance of $\sim$2.3 $R_\odot$ where the magnetic field lines extend mainly in a radial direction and the type II radio emission ceased. Beyond $\sim$2 $R_\odot$ the shock-to-field geometry becomes predominantly quasi-parallel ($\theta_{Bn}$ $<$ 45$^\circ$) in which case diffusive shock acceleration (DSA), also referred to as the first order Fermi process, is the dominant particle acceleration mechanism. In this mechanism, particles are accelerated by successive reflections between the shock down and upstream regions due to the presence of magnetic turbulence. According to DSA theory, the rate of electron acceleration and, hence, the maximum energy attained, is dependent on the shock-to-field geometry, as well as the shock strength \citep{Jokipii1987}. The electron acceleration efficiency would be considerably lower in the quasi-parallel configuration due to the long times required to energize particles (they spend most of their time random walking in the upstream or downstream regions; \citealt{Guo2010,Verkhoglyadova2015}). Given that the shock nose becomes quasi-parallel, the electron acceleration efficiency may have reduced at this point and may be an explanation of the cessation of the radio emission despite an increasing $M_A$. This suggests that shock geometry is an important factor in the acceleration of electrons and the presence of radio emission similar to results of \cite{Kozarev2015}, \cite{SalasMatamoros2016} and \cite{Zucca2018}.

\section{Conclusions}
\label{conclusions}
We present a study of the formation and evolution of a CME-driven shock using three commonly used methods, namely: shock geometry, a comparison of CME speed to a model of the coronal Alfvén speed, and the type II band-splitting method. We were able to determine $M_A$ in the corona at heliocentric distance range of $\sim$1.4 to $\sim$3 $R_\odot$. The results from all three methods were initially similar ($\sim$1.5) but diverge later on. The divergence may be a result of the inherent uncertainties associated with each method. As a general trend, the results suggest $M_A$ was initially $\sim$1.5 and increased up to 4 over a time frame $\sim$17 minutes. Type II radio emission, coming from the nose region of the CME, began when the shock reached a heliocentric distance of $\sim$1.6 $R_\odot$ and $M_A$ was in the range 1.4-2.4. Despite an increasing $M_A$ (up to 4), the emission ceased when the shock front reached $\sim$2.4 $R_\odot$. We suggest this is a result of a change in shock geometry, that is, the shock was no longer quasi-perpendicular and efficient electron acceleration and radio emission was inhibited. These results provide insight into the shock conditions necessary for producing type II emission suggesting a supercritical $M_A$ and favorable shock geometry is required for the acceleration of energetic electrons.\newline \indent 
The focus of future works will be to estimate the shock-to-field angle by fitting a spherical geometric surface to the shock using shock wave kinematics and combining it with a PFSS model. This will allow us to validate our assumption, determine the region of electron acceleration along the shock front in a 3D context and give quantitative properties of shock geometry necessary for type II emission. 
\newline
\newline
\newline
\noindent
\textit{Acknowledgments and Data Availability} \newline
C. A. M. is supported by an Irish Research Council Government of Ireland Postgraduate Scholarship. E. P. C. is supported by the European Commission Horizon 2020 INFRADEV-1-2017 LOFAR4SW project, no. 777442. We would like to acknowledge NOAA’s National Centers for Environmental Information SUVI team for providing data on the 2017 September 2 event, particular thanks to Dan Seaton for preparing the data. We are grateful to the GOES, Wind and LASCO teams for open access to their data. I-LOFAR received funding from Science Foundation Ireland (SFI) grant no. 15/RI/3204. The I-LOFAR data can be obtained from http://data.lofar.ie or on request to observer@lofar.ie. ORFÉES is part of the FEDOME project, partly funded by the French Ministry of Defense. We would like to thank Pietro Zucca for his assistance with the \cite{Zucca2014} model. 

\bibliographystyle{aa}
\bibliography{references.bib}

\begin{thebibliography}{52}
\expandafter\ifx\csname natexlab\endcsname\relax\def\natexlab#1{#1}\fi

\bibitem[{Berger {et~al.}(2012)Berger, Liu, \& Low}]{Berger2012}
Berger, T.~E., Liu, W., \& Low, B.~C. 2012, The Astrophysical Journal, 758, L37

\bibitem[{Bougeret {et~al.}(1995)Bougeret, Kaiser, Kellogg, Manning, Goetz,
  Monson, Monge, Friel, Meetre, Perche, Sitruk, \& Hoang}]{Bougeret1995}
Bougeret, J., Kaiser, M., Kellogg, P., {et~al.} 1995, Space Science Reviews,
  71, 231

\bibitem[{Brueckner {et~al.}(1995)Brueckner, Howard, Koomen, Korendyke,
  Michels, Moses, Socker, Dere, Lamy, Llebaria, Bout, Schwenn, Simnett,
  Bedford, \& Eyles}]{Brueckner1995}
Brueckner, G.~E., Howard, R.~A., Koomen, M.~J., {et~al.} 1995, Solar Physics,
  162, 357

\bibitem[{Carley {et~al.}(2013)Carley, Long, Byrne, Zucca, Shaun~Bloomfield,
  McCauley, \& Gallagher}]{Carley2013}
Carley, E.~P., Long, D.~M., Byrne, J.~P., {et~al.} 2013, Nature Physics, 9, 811

\bibitem[{Cho {et~al.}(2007)Cho, Gary, Lee, Moon, \& Park}]{Cho2007}
Cho, K.-S., Gary, D.~E., Lee, J., Moon, Y.-J., \& Park, Y.~D. 2007, The
  Astrophysical Journal, 665, 799

\bibitem[{Chrysaphi {et~al.}(2018)Chrysaphi, Kontar, Holman, \&
  Temmer}]{Chrysaphi2018}
Chrysaphi, N., Kontar, E.~P., Holman, G.~D., \& Temmer, M. 2018, The
  Astrophysical Journal, 868, 10

\bibitem[{De~Hoffmann \& Teller(1950)}]{Hoffmann1950}
De~Hoffmann, F. \& Teller, E. 1950, Physical Review, 80, 692

\bibitem[{Du {et~al.}(2015)Du, Kong, Chen, Feng, Wang, \& Li}]{Du2015}
Du, G., Kong, X., Chen, Y., {et~al.} 2015, 1

\bibitem[{Edmiston \& Kennel(1984)}]{Edmiston1984}
Edmiston, J.~P. \& Kennel, C.~F. 1984, Journal of Plasma Physics, 32, 429

\bibitem[{Farris {et~al.}(1991)Farris, Petrinec, \& Russell}]{Farris1991}
Farris, M.~H., Petrinec, S.~M., \& Russell, C.~T. 1991, Geophysical Research
  Letters, 18, 1821

\bibitem[{Frassati {et~al.}(2019)Frassati, Susino, Mancuso, \&
  Bemporad}]{Frassati2019}
Frassati, F., Susino, R., Mancuso, S., \& Bemporad, A. 2019, The Astrophysical
  Journal, 871, 212

\bibitem[{Gary(2001)}]{Gary2001}
Gary, G.~A. 2001, Solar Physics, 203, 71

\bibitem[{Ginzburg \& Zhelezniakov(1993)}]{Ginzburg1993}
Ginzburg, V.~L. \& Zhelezniakov, V.~V. 1993, Soviet Astronomy, Vol. 2, p.653,
  2, 653

\bibitem[{Gopalswamy {et~al.}(2011)Gopalswamy, Nitta, Akiyama,
  M{\"{a}}kel{\"{a}}, \& Yashiro}]{Gopalswamy2011b}
Gopalswamy, N., Nitta, N., Akiyama, S., M{\"{a}}kel{\"{a}}, P., \& Yashiro, S.
  2011, 72, 0

\bibitem[{Gopalswamy \& Yashiro(2011)}]{GopalswamyYash2011}
Gopalswamy, N. \& Yashiro, S. 2011, The Astrophysical Journal, 736, L17

\bibitem[{Grechnev {et~al.}(2011)Grechnev, Afanasyev, Uralov, Chertok,
  Eselevich, Eselevich, Rudenko, \& Kubo}]{Grechnev2011a}
Grechnev, V.~V., Afanasyev, A.~N., Uralov, A.~M., {et~al.} 2011, Solar Physics,
  273, 461

\bibitem[{Guo \& Giacalone(2010)}]{Guo2010}
Guo, F. \& Giacalone, J. 2010, The Astrophysical Journal, 715, 406

\bibitem[{Holman \& Pesses(1983)}]{Holman1983}
Holman, G.~D. \& Pesses, M.~E. 1983, The Astrophysical Journal, 267, 837

\bibitem[{Jokipii(1987)}]{Jokipii1987}
Jokipii, J.~R. 1987, The Astronomical Journal, 313, 842

\bibitem[{Kouloumvakos {et~al.}(2014)Kouloumvakos, Patsourakos, Hillaris,
  Vourlidas, Preka-Papadema, Moussas, Caroubalos, Tsitsipis, \&
  Kontogeorgos}]{Kouloumvakos2014}
Kouloumvakos, A., Patsourakos, S., Hillaris, A., {et~al.} 2014, Solar Physics,
  289, 2123

\bibitem[{Kozarev {et~al.}(2011)Kozarev, Korreck, Lobzin, Weber, \&
  Schwadron}]{Kozarev2011}
Kozarev, K.~A., Korreck, K.~E., Lobzin, V.~V., Weber, M.~A., \& Schwadron,
  N.~A. 2011, The Astrophysical Journal, 733, L25

\bibitem[{Kozarev {et~al.}(2015)Kozarev, Raymond, Lobzin, \&
  Hammer}]{Kozarev2015}
Kozarev, K.~A., Raymond, J.~C., Lobzin, V.~V., \& Hammer, M. 2015, The
  Astrophysical Journal, 799, 167

\bibitem[{Ma {et~al.}(2011)Ma, Raymond, Golub, Lin, Chen, Grigis, Testa, \&
  Long}]{Ma2011}
Ma, S., Raymond, J.~C., Golub, L., {et~al.} 2011, Astrophysical Journal, 738,
  160

\bibitem[{Mahrous {et~al.}(2018)Mahrous, Alielden, Vr{\v{s}}nak, \&
  Youssef}]{Mahrous2018}
Mahrous, A., Alielden, K., Vr{\v{s}}nak, B., \& Youssef, M. 2018, Journal of
  Atmospheric and Solar-Terrestrial Physics, 172, 75

\bibitem[{Maloney \& Gallagher(2011)}]{Maloney2011}
Maloney, S.~A. \& Gallagher, P.~T. 2011, The Astrophysical Journal Letters,
  736, 5

\bibitem[{Mancuso {et~al.}(2019)Mancuso, Frassati, Bemporad, \&
  Barghini}]{Mancuso2019}
Mancuso, S., Frassati, F., Bemporad, A., \& Barghini, D. 2019, A{\&}A, 624, 2

\bibitem[{Mann {et~al.}(2002)Mann, Classen, Keppler, \& Roelof}]{Mann2002}
Mann, G., Classen, H.~T., Keppler, E., \& Roelof, E.~C. 2002, Astronomy {\&}
  Astrophysics, 391, 749

\bibitem[{Mann {et~al.}(1995)Mann, Classen, \& Aurass}]{Mann1995}
Mann, G., Classen, T., \& Aurass, H. 1995, Astronomy and Astrophysics, 295, 775

\bibitem[{Melrose(1975)}]{Melrose1975}
Melrose, D.~B. 1975, Solar Physics, 43, 79

\bibitem[{Morosan {et~al.}(2019)Morosan, Carley, Hayes, Murray, Zucca, Fallows,
  McCauley, Kilpua, Mann, Vocks, \& Gallagher}]{Morosan2019}
Morosan, D.~E., Carley, E.~P., Hayes, L.~A., {et~al.} 2019, Nature Astronomy, 1

\bibitem[{Nelson \& Melrose(1985)}]{Nelson1985}
Nelson, G.~J. \& Melrose, D.~B. 1985, in Solar radiophysics: Studies of
  emission from the sun at metre wavelengths, 333--359

\bibitem[{Poomvises {et~al.}(2012)Poomvises, Gopalswamy, Yashiro, Kwon, \&
  Olmedo}]{Poomvises2012}
Poomvises, W., Gopalswamy, N., Yashiro, S., Kwon, R.~Y., \& Olmedo, O. 2012,
  Astrophysical Journal, 758, 1

\bibitem[{Rouillard {et~al.}(2016)Rouillard, Plotnikov, Pinto, Tirole, Lavarra,
  Zucca, Vainio, Tylka, Vourlidas, Rosa, Linker, Warmuth, Mann, Cohen, \&
  Mewaldt}]{Rouillard_2017}
Rouillard, A.~P., Plotnikov, I., Pinto, R.~F., {et~al.} 2016, The Astrophysical
  Journal, 833, 45

\bibitem[{Russell \& Mulligan(2002)}]{RussellMulligan2002}
Russell, C.~T. \& Mulligan, T. 2002, Planetary and Space Science, 50, 527

\bibitem[{Salas-Matamoros {et~al.}(2016)Salas-Matamoros, Klein, \&
  Rouillard}]{SalasMatamoros2016}
Salas-Matamoros, C., Klein, K.-L., \& Rouillard, A.~P. 2016, Astronomy {\&}
  Astrophysics, 590, A135

\bibitem[{Schmidt \& Cairns(2012)}]{Schmidt2012}
Schmidt, J.~M. \& Cairns, I.~H. 2012, Journal of Geophysical Research: Space
  Physics, 117, n/a

\bibitem[{Seaton \& Darnel(2018)}]{Seaton2018}
Seaton, D.~B. \& Darnel, J.~M. 2018, The Astrophysical Journal, 852, L9

\bibitem[{Seiff(1962)}]{Seiff1962}
Seiff, A. 1962, in Gas Dynamics in Space Explorations: Recent Information on
  Hypersonic Flow Fields, Vol.~24, 19

\bibitem[{Smerd {et~al.}(1974)Smerd, Sheridan, \& Stewart}]{Smerd1974}
Smerd, S.~F., Sheridan, K.~V., \& Stewart, R.~T. 1974, Symposium -
  International Astronomical Union, 57, 389

\bibitem[{Spreiter {et~al.}(1966)Spreiter, Summers, \& Alksne}]{Spreiter1966}
Spreiter, J.~R., Summers, A.~L., \& Alksne, A.~Y. 1966, Planetary and Space
  Science, 14, 223

\bibitem[{Stansby(2019)}]{Stansby2019}
Stansby, D. 2019, {dstansby/pfsspy: pfsspy 0.1.2}

\bibitem[{Street {et~al.}(1994)Street, Ball, \& Melrose}]{Street1994}
Street, A.~G., Ball, L., \& Melrose, D.~B. 1994, Publications of the
  Astronomical Society of Australia, 11, 21

\bibitem[{Verkhoglyadova {et~al.}(2015)Verkhoglyadova, Zank, \&
  Li}]{Verkhoglyadova2015}
Verkhoglyadova, O.~P., Zank, G.~P., \& Li, G. 2015, Physics Reports, 557, 1

\bibitem[{Vourlidas \& Bemporad(2012)}]{Vourlidas2012}
Vourlidas, A. \& Bemporad, A. 2012, AIP Conference Proceedings, 1436, 279

\bibitem[{Vr{\v{s}}nak {et~al.}(2001)Vr{\v{s}}nak, Aurass, Magdaleni{\'{c}}, \&
  Gopalswamy}]{Vrsnak2001}
Vr{\v{s}}nak, B., Aurass, H., Magdaleni{\'{c}}, J., \& Gopalswamy, N. 2001,
  Astronomy {\&} Astrophysics, 377, 321

\bibitem[{Vr{\v{s}}nak {et~al.}(2002)Vr{\v{s}}nak, Magdalenic, Aurass, \&
  Mann}]{Vrsnak_Mag2002}
Vr{\v{s}}nak, B., Magdalenic, J., Aurass, H., \& Mann, G. 2002, A{\&}A, 396,
  673

\bibitem[{Wild(1962)}]{Wild1962}
Wild, J.~P. 1962, Journal of the Physical Society of Japan Supplement, 17, 249

\bibitem[{Wild \& McCready(1950)}]{Wild1950}
Wild, J.~P. \& McCready, L.~L. 1950, Australian Journal of Scientific Research
  A, vol. 3, p.387, 3, 387

\bibitem[{Zimovets {et~al.}(2012)Zimovets, Vilmer, Chian, Sharykin, \&
  Struminsky}]{Zimovets2012}
Zimovets, I., Vilmer, N., Chian, A. C.~L., Sharykin, I., \& Struminsky, A.
  2012, Astronomy {\&} Astrophysics, Volume 547, id.A6 13 pp., 547

\bibitem[{Zucca {et~al.}(2014{\natexlab{a}})Zucca, Carley, Bloomfield, \&
  Gallagher}]{Zucca2014}
Zucca, P., Carley, E.~P., Bloomfield, D.~S., \& Gallagher, P.~T.
  2014{\natexlab{a}}

\bibitem[{Zucca {et~al.}(2018)Zucca, Morosan, Rouillard, Fallows, Gallagher,
  Magdalenic, Klein, Mann, Vocks, Carley, Bisi, Kontar, Rothkaehl, Dabrowski,
  Krankowski, Anderson, Asgekar, Bell, Bentum, Best, Blaauw, Breitling,
  Broderick, Brouw, Br{\"{u}}ggen, Butcher, Ciardi, Geus, Deller, Duscha,
  Eisl{\"{o}}ffel, Garrett, Grie{\ss}meier, Gunst, Heald, Hoeft,
  H{\"{o}}randel, Iacobelli, Juette, Karastergiou, Leeuwen, McKay-Bukowski,
  Mulder, Munk, Nelles, Orru, Paas, Pandey, Pekal, Pizzo, Polatidis, Reich,
  Rowlinson, Schwarz, Shulevski, Sluman, Smirnov, Sobey, Soida, Thoudam,
  Toribio, Vermeulen, van Weeren, Wucknitz, \& Zarka}]{Zucca2018}
Zucca, P., Morosan, D.~E., Rouillard, A.~P., {et~al.} 2018, Astronomy {\&}
  Astrophysics, 615, A89

\bibitem[{Zucca {et~al.}(2014{\natexlab{b}})Zucca, Pick, Demoulin, Kerdraon,
  Lecacheux, \& Gallagher}]{Zucca2014b}
Zucca, P., Pick, M., Demoulin, P., {et~al.} 2014{\natexlab{b}}

\end{thebibliography}
\end{document}